\documentclass[runningheads,citeauthoryear]{apinv2021}
\usepackage{amsmath}
\usepackage{epsfig,cite,graphics}
\usepackage{graphicx}
\usepackage{caption} 
\usepackage{marvosym} 
\usepackage[utf8]{inputenc}
\usepackage{multirow}

\begin{document}

\title{Unveiling the multiple periodicities of the cataclysmic variable LS~Cam}
\titlerunning{ Periodicities of LS~Cam}
\author{S. Y. Stefanov$^{1,2}$}
\authorrunning{Stefanov, S.}
\tocauthor{S. Y. Stefanov} 

\institute{Institute of Astronomy and National Astronomical Observatory, Bulgarian Academy of Sciences, Tsarigradsko Shose 72, 
BG-1784 Sofia, Bulgaria  \\
	\and  Department of Astronomy, Sofia University "St. Kliment Ohridski", James Bourchier 5, BG-1164 Sofia, Bulgaria \\
	\email{sjonkov@uni-sofia.bg}      }

\papertype{Submitted on 19 May 2021; Accepted on -}	
\maketitle

\begin{abstract}
$TESS$ photometric data of LS~Cam from sectors 19, 20 and 26 are analysed. The obtained power spectra from sectors 19 and 20 show multiple periodicities - orbital variations ($P_{orb} = 0.14237$ days), slightly fluctuating superorbital variation ($ P_{so} \approx 4.03$ days) and permanent negative superhump ($P_{-sh} = 0.1375$ days). In sector 26 an additional positive superhump ($P_{+sh} = 0.155$ days) is present. Using relations from literature, the mass ratio and the masses of the two components are estimated to be $q =0.24$, $M_1 = 1.26M_{\sun}$, and $M_2 = 0.30 M_{\sun}$ respectively.

\end{abstract}

\section{Introduction}
Cataclysmic variables (CVs) are binary systems consisting of a white dwarf primary and a Roche-lobe filling late-type star secondary. 
Matter from the secondary is usually transferred through the L$_1$ Lagrange point to the primary via an accretion disc. 
In the presence of strong magnetic fields generated by the primary, the matter is accreted along the magnetic lines onto the magnetic poles. 
A detailed overview of cataclysmic systems is presented in Warner (2003) and Hellier (2001). 
Some CVs display quasi-periodic changes in brightness with periods slightly shorter or slightly 
longer than the orbital period (P$_{orb}$) of the system. These phenomena are called superhumps. 
They can be either positive/apsidal $(P_{+sh})$  - a few per cent longer than P$_{orb}$, 
or a few per cent shorter - negative/nodal $(P_{-sh})$. Superhumps are believed to be a beat period of P$_{orb}$ 
and the apsidal or nodal motion of the disk, but their exact cause remains unknown. 
If the superhump period remains stable for many thousands of orbital cycles it is called permanent superhump 
(e.g. Patterson 1999). In some cases along with a negative superhump, 
a superorbital signal $(P_{so})$ with a significantly longer period than P$_{orb}$ is detected. 

\begin{figure}[h]
\includegraphics[width=\columnwidth]{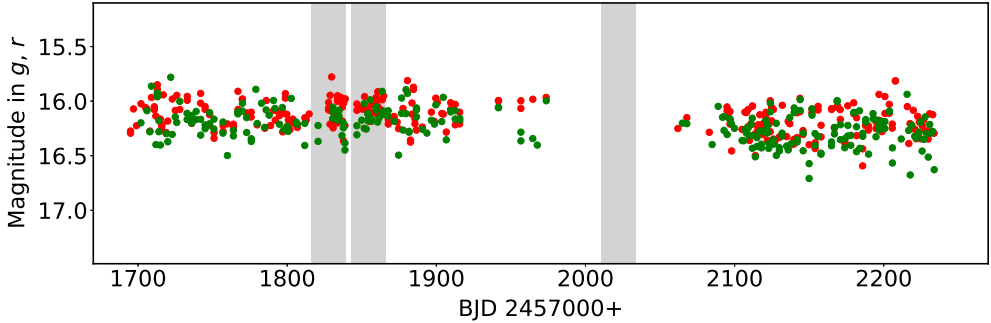}
\caption{ZTF light curves in $g$ and $r$ are shown with green and red respectively. The photometry errors are smaller than the symbol size. The shaded regions represent the times of $TESS$ sectors 19, 20, and 26 in which LS~Cam was observed. }
\label{fig:ZTF}
\end{figure}

LS~Cam is a cataclysmic variable that was first classified as a quasar candidate with 
the identifier~HS~0551+7241 in the Hamburg-CfA Quasar Survey (Engels et al. 1998). 
Later, Dobrzycka et al. (1998) show the CV nature of HS 0551+7241 and study its behavior using 
spectroscopic and photometric observations. Their analysis of the spectroscopic data shows evidence of
short $\sim$ 50 min fluctuations superimposed on longer $\sim$ 4-hour variations. 
They also consider the presence of eclipses and suggest that the binary is a candidate 
intermediate polar based on a relatively strong HeII $\lambda$4686 emission line with short fluctuations. 
The long-term photometric behavior is also studied in Dobrzycka et al. (1998). 
The average historic magnitude of LS~Cam is 16.8~mag in $B$~band, except for a 2 year 
long low state starting in January 1994, when the system's brightness dropped by $\Delta B \sim 2.5$ mag. 
Data from the Zwicky Transient Facility (ZTF; Masci et al. 2019) in the period of 2019 to 2021 show 
LS~Cam in a stable high state with a mean magnitude of 16.24 in $g$ band and 16.17 in $r$ band (Fig.~\ref{fig:ZTF}). 

Other mentions of LS~Cam in the literature include a work of Rodr{\'\i}guez-Gil (2005), where the binary is classified as a non-eclipsing SW~Sex type system and a work of Thorstensen~et~al.~(2017), where an improved orbital period $P_{orb} = 0.1423853(5)$ days $\approx$ 3.42 hours is given.

\section{Observations and results}
Photometry of LS~Cam with 120 seconds cadence was obtained from sectors 19, 20, and 26 of the Transiting Exoplanet Survey Satellite ($TESS$; Ricker et al. 2015). Each sector includes $\sim$ 24 days of continuous photometry, with one day pause during orbit perigee passage. The data are publicly available in the Mikulski Archive for Space Telescopes (MAST)\footnote{http://archive.stsci.edu/}. More information about the sectors of the $TESS$ telescope can be found at the $TESS$ website\footnote{https://tess.mit.edu/observations/}. 
The data are reduced with the SPOC pipeline (Jenkins et al. 2016) and from each sector is subtracted its mean flux level. Sectors 19 and 20 are analyzed together, separately from sector 26. Periodogram analysis of the two sets of data was performed using the Lomb-Scargle (LS) method (Lomb 1976; Scargle 1982) with an Astropy routine (Astropy Collaboration et al. 2013, 2018). To estimate the uncertainties of the obtained periods from the power spectra, a bootstrap method with sample size equal to half of the data is used.

\begin{figure}[h!]
\begin{center}
\includegraphics[scale=0.23]{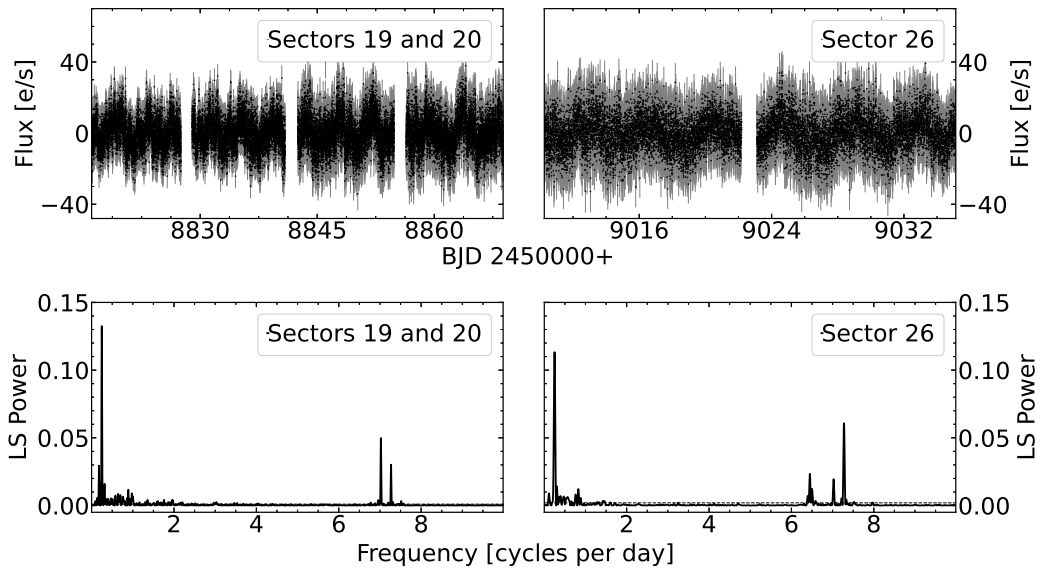}
\caption{The two sets of $TESS$ photometry (black dots) and errors (gray lines). The corresponding periodograms are below each light curve.}
\label{fig:periodograms}
\end{center}
\end{figure}

The obtained power spectra contain four significant periodicities - the orbital period $(P_{orb})$, a superorbital variation $(P_{so})$, a permanent negative superhump $(P_{-sh})$ and in sector 26 only - a positive superhump $(P_{+sh})$. The values of all obtained periods, duration, start, and sectors of each of the two data sets are shown in Table~\ref{tab:Results}. The uncertainty is expressed in terms of the least significant digit in brackets after the measured period value. Figure~\ref{fig:periodograms} shows all photometric data from the $TESS$ sectors and the corresponding periodograms. A preliminary search for periodicities was performed in the frequency window between 0.001 and 300 cycles per day. All significant periodic signals are located in the frequency window between 0.01 and 10 cycles per day, shown in the power spectra in the bottom two panels of Figure~\ref{fig:periodograms}.

\vspace{-0.3cm}
\renewcommand{\arraystretch}{1.5}
\setlength{\tabcolsep}{5pt}
\begin{table}[]
    \centering
     \begin{tabular}{ccccccc}\\ 
\hline
\multirow{2}{*}{Sector} & start & Duration & P$_{orb}$ & P$_{so}$ & P$_{-sh}$ & P$_{+sh}$ \\
 &JD 2450000+& [days] & [days] & [days] & [days] & [days] \\
\hline
  19 & 8816.08 & 24.10 & \multirow{2}{*}{0.1424(3)}& \multirow{2}{*}{4.053(3)}& \multirow{2}{*}{0.1375(4)}& \multirow{2}{*}{-}\\
\vspace{5pt}
20 & 8842.50 & 24.79 & & & & \\
26 & 9010.26 & 23.95 & 0.14237(2) & 4.006(9) & 0.137(1) & 0.155(1)\\
\hline
\\\\
    \end{tabular}
    \caption{Results from periodogram analysis of the $TESS$ observations.}
    \label{tab:Results}
\end{table}

\section{Discussion}
The beat period of $P_{orb}$ and $P_{-sh}$ is:
\begin{equation}
    \dfrac{1}{P_{-sh}} - \dfrac{1}{P_{orb}} = \dfrac{1}{P_{so}}
    \label{eq.beat}
\end{equation}
The expected $P_{so}$ calculated with the obtained $P_{orb}$ and $P_{-sh}$ is 4.00$\pm$0.33 days. This value is in agreement with the $P_{so}$ values in Table~\ref{tab:Results}, implying that the negative superhump is a beat period of the orbital period and the disk precession motion.

The presence of simultaneous negative and positive superhumps would suggest that the accretion disc of LS~Cam is undergoing both apsidal and nodal precession motions. The existence of both periodicities have been observed in a few other systems - TT~Ari (Belova et al. 2013, Kraicheva et al. 1999), AQ~Men (Iłkiewicz et al. 2021), and V603 Aql (Patterson 1997).

Suleimanov et al. (2004) explore the transition between negative and positive superhumps in V603~Aql. In this work, they argue that such transitions happen due to change in the accretion rate in the system. The authors suggest that that the disk inclination increases with the distance from the primary. The farther a ring of matter is from the white dwarf, the higher its inclination to the orbital plane. This effect is stronger when the disk is larger, thus when the accretion rate lowers, this effect becomes smaller and the apsidal motion of the disc becomes significant enough to appear in the photometric variability of the system. This can explain the behaviour of LS~Cam - the mean flux of sectors 19 and 20 is 43.8 e/s. For sector 26 it is 26.1 e/s, implying that the disk is smaller and the accretion rate is lower.

An estimation of some system parameters of LS~Cam can be obtained using the positive superhump excess $\epsilon_{+} = (P_{+sh} - P_{orb})/P_{orb}$ and the negative superhump deficit $\epsilon_{-} = (P_{-sh} - P_{orb})/P_{orb}$. With the periods given in Table~1 and $P_{orb} = 0.1423853 \pm 0.0000005$ days (Thorstensen et al. 2017), the values of $\epsilon_{+}$ and $\epsilon_{-}$ for LS~Cam are $0.088 \pm 0.007$ and $0.034 \pm 0.007$ respectively. The excess and deficit are shown to correlate with the mass ratio $(q)$ and the orbital period of the system (e.g. Retter et al. 2002; Patterson et al. 2005). Wood et al. (2009) give a relation between $q$ and $\epsilon_{-}$ derived from particle simulations: 

\begin{equation}
      q = -0.192|\epsilon_{-}|^{1/2} + 10.37|\epsilon_{-}| - 99.83|\epsilon_{-}|^{3/2} + 451.1|\epsilon_{-}|^2.
      \label{eq:2}
\end{equation}

They also provide a relation between $q$ and $\epsilon_{+}$:

\begin{equation}
q = 3.733\epsilon_{+} - 7.898\epsilon_{+}^2,
\label{eq:3}
\end{equation}

Using the obtained values of $\epsilon_{+}$ and $\epsilon_{-}$, the resulting value of $q$ in Eq.~\ref{eq:2} is $q = 0.21 \pm 0.09$, and using Eq.~\ref{eq:3} is $q = 0.26 \pm 0.02$. The uncertainty here is estimated using only the errors from the periodogram analysis.

From the value of $q = 0.24 \pm 0.02$ and the mean empirical mass-period relationship $M_2 = 0.065[P_{orb}(hr)]^{1.25} $ (Warner 2003), the masses of the two components are estimated to be $M_1 = 1.26 \pm 0.10M_{\sun}$ and $M_2 =0.302 \pm 0.01 M_{\sun}$. The values for $q$, $M_1$ and $M_2$ estimated in this work are similar to the rough estimations done by Dobrzycka et al. (1998).

\vspace{1cm}

{\bf Conclusions: }
In this work analysis of three sectors of $TESS$ photometry of the cataclysmic variable LS~Cam is performed. The Power spectra show the existence of orbital variations with $P_{orb} = 0.14237 \pm 0.00002$ days, slightly fluctuating superorbital variation  with $ P_{so} \approx 4.03$ days and permanent negative superhump $P_{-sh} = 0.1375 \pm 0.0004$ days. An additional positive superhump with $P_{+sh} = 0.155 \pm 0.001$ days existing simultaneously with the other three variations is present in the sector 26. 

Using the relations between $q$, $\epsilon_{+}$, $\epsilon_{-}$ and  the  mass-period relationship, several system parameters are estimated. The values for the mass ratio $q = 0.24 \pm 0.02$, and mass of the components are $M_1 = 1.26 \pm 0.10M_{\sun}$ and  $M_2 =0.302 \pm 0.01 M_{\sun}$.

\vspace{1cm}

{\small {\bf Acknowledgments: }
The author is grateful to prof.~R.~Zamanov for the helpful bits of advice and fruitful discussions. 
This work was supported by the  Bulgarian National Science Fund project  number K$\Pi$-06-H28/2 08.12.2018
"Binary stars with compact object".}

\end{document}